\documentclass[manuscript=article]{achemso}
\usepackage[utf8]{inputenc}
\usepackage[intlimits]{amsmath}
\usepackage{amssymb}
\usepackage{xcolor}
\usepackage{hyperref}
\usepackage[symbol]{footmisc}
\setkeys{acs}{
  abbreviations = false,
  articletitle  = true,
  biblabel      = brackets,
  biochem       = false,
  doi           = false,
  etalmode      = firstonly,
  keywords      = true,
  maxauthors    = 15,
  super         = true
}
\author{Marsel Karmo}
\affiliation{Institut für Physik, Technische Universität Ilmenau, 98693 Ilmenau, Germany}
\author{Hartmut Grille}
\affiliation{Institut für Physik, Technische Universität Ilmenau, 98693 Ilmenau, Germany}
\author{Isaac Azahel Ruiz Alvarado}
\affiliation{Lehrstuhl für Theoretische Materialphysik, Universität Paderborn, 33098 Paderborn, Germany}
\email{karmo@mail.uni-paderborn.de}
\title{Calculating the Stability of Different Surfaces of $\mathbf{GaAs_xP_{1-x}}$ Mixed-Crystals using the Virtual Crystal Approximation}
\date{}

\begin{document}
 \begin{abstract}
The theoretical treatment of mixed-crystals is very demanding. A straight-forward approach to attack this problem is using a super cell method (SCM). Another one is the Virtual Crystal Approximation (VCA), which is a feature of the Vienna Ab initio Simulation Package (VASP). For comparison we use both methods to calculate the total energy ($E_{tot}$) and the density of states (DOS) of bulk $GaAs_xP_{1-x}$. We then apply VCA to compute the stability of different surfaces using an extended version of the surface formation energy $\Omega$. Our calculations show, on one hand, a working VCA implementation with its flaws (overestimation of $E_{tot}$) and strengths (well modelling of DOS). On other hand, a further result is that bulk of the slab of a mixed-crystal has a minor influence on the configuration of the surface.
 \end{abstract}
\section*{INTRODUCTION}
The efforts to combine different optical active materials into  high-efficient multi-junction solar cells have a long history \cite{yamaguchi2001super}. Combining various thin films is  challenging because of strain due to different lattice constants, different chemistry, the difficulties of contacting, and more \cite{almosni2018material}. For example, the direct deposition of $GaAs$ on $Si$ (as a substrate) leads to strain because of lattice mismatch. One method to avoid this is to use the material system $GaAs_{x}P_{1-x}$ as a buffer layer\cite{mei1997liquid,grassman2010characterization,prete2023lattice}. Starting with a $GaP$ layer (with nearly the same lattice constant as $Si$) one subsequently deposits $GaAs_{x}P_{1-x}$ layers with increasing $x$ until, at least, the mixed-crystal becomes a direct semiconductor \cite{gharibshahian2021modeling}. 
\\
Another point of interest is the surface geometry of solar cell materials which may affect adsorption of other atoms (molecules) which could change the resistance and efficiency of the solar cell. For pure $GaP$ and $GaAs$ surfaces, theoretical \cite{hahn2003p,khosroabadi2018high,karmo2022reconstructions} and experimental\cite{ohtake2008surface,yoshikawa1996surface,frisch19992,freundlich2010single} investigations have been performed in the past. But little is known about the surfaces of the mixed crystal $GaAs_{x}P_{1-x}$.
\\
The dependence of the pure $GaP$ and $GaAs$ crystals surface reconstruction on the control process (growth, cooling, tempering, etc.) are well known from experimental findings.\cite{ohtake2008surface,law2003reflectance,shu2005incorporation}.
The most common surface reconstructions reported in experiments are the $(2 \times 2)$-2D-2H for $GaP$ ($P$-rich) and the $\beta2(2\times4)$ for $GaAs$ ($As$-rich) within the $(2\times4)$-range \cite{supplie2017situ,ohtake2002atomic,mccoy1998determination}, which are in good agreement with theoretical calculations \cite{hahn2003p,karmo2022reconstructions}.
\\
Investigating the surface of a $GaAs_{x}P_{1-x}$ mixed-crystal one would, naively, expect a more or less sharp transition from one geometry to the other at a certain concentration $x$. On the other hand, the composition of the surface depends on the details of the cooling process (temperature, open or closed precursors or the ambience in general) \cite{guan2016enhanced,doscher2017gap,supplie2017situ,hannappel2004apparatus}. Usually the cool down of the sample is done in such a way that a closed P-surface forms. It is reported that the geometry of this type of surface is $2\times2$/$2\times1$ \cite{supplie2020quantification,supplie2018optical}.
\\
To compare two given surfaces energetically, it is common practice to calculate the surface formation energy $\Omega$ \cite{bechstedt2012principles}. We apply an extended version of this quantity to crystals with compositional disorder.
\\
The properties of a crystal with compositional disorder can only be calculated approximately. One option is a super-cell method (SCM), i.e. one takes a large elementary cell with as many atoms as possible (limited primarily by the hardware available) and hopes for self-averaging. Another approach is to execute the average analytically in the approximation creating a substitute system. This is the case for the Virtual Crystal Approximation (VCA)\cite{schoen1969augmented,bellaiche2000virtual,cai2021lattice}, setting virtual anions of each type weighted with the composition $x$ on each anion site. Not only does this overcome the need of circumventing the periodic boundary condition with a large super-cell but  one can also model rather complicated structures (step-graded buffer \cite{zakaria2012comparison}, quantum wells\cite{citation-0}, etc.). We use the Vienna Ab initio Simulation Package (VASP) \cite{VASPGoog11:online} which offers an implementation of VCA.
\\
While a bulk mixed-crystal $GaP/GaAs_{x}P_{1-x}$ has a rather simple and high symmetric lattice geometry, it is possible to apply both methods without hitting the limits of the computer resources. On the other hand, in VASP a crystal with a surface is described by a super lattice (crystal-vacuum). This structure is not only rather complicated but also requires a large number of atoms to approximate the semi-infinite slab. This and the size of the vacuum are also dimensioned to minimize super-lattice effects, i.e. interaction with the other crystal layers. A SCM for such a structure is computational too demanding -- the desire for self-averaging and the three-dimensionality pushes the necessary number of atoms easily above a few thousands. Therefore we investigated the surfaces of  $GaAs_{x}P_{1-x}$ crystals only with VCA.
However, we compare the DOS for $x=0.5$ with that of a random configuration without any averaging.
\section*{METHODOLOGY}
The VCA method was originally developed by Bellaiche and Vanderbilt \cite{bellaiche2000virtual} based on the weighted averaging of the pseudo-potentials. We describe the method briefly.
The Hamilton operator of a crystal is given by:
\begin{eqnarray}\label{eq:1}
\widehat{H} =-\sum_{i}\frac{\hbar^2}{2m}
\overrightarrow{\nabla_{i}}^{2}+\sum_{i\neq j}V_{i}(\overrightarrow{r_{i}},\overrightarrow{r_{j}})
\end{eqnarray}
with a translational invariant potential (with a lattice vector $\overrightarrow{R}$) :
\begin{eqnarray}
V_{i}(\overrightarrow{r_{i}}+\overrightarrow{R},\overrightarrow{r_{j}}+\overrightarrow{R})=V_{i}(\overrightarrow{r_{i}},\overrightarrow{r_{j}})
\end{eqnarray}
A mixed-crystal lacks this invariance. The simplest VCA method reintroduces the translational invariance by replacing the "disordered" potential with an ordered, averaged potential (in our case):
\begin{eqnarray}
V_{VCA}(\overrightarrow{r_{i}},\overrightarrow{r_{j}})=xV_{GaAs}(\overrightarrow{r_{i}},\overrightarrow{r_{j}})+(1-x)V_{GaP}(\overrightarrow{r_{i}},\overrightarrow{r_{j}})
\end{eqnarray}
To give a picture, one can imagine the substitute system as an ordered crystal with virtual $As/P$ anions weighted by concentration $x$ ($As$) and $1-x$ ($P$) on each anion place. 
\\
Firstly, we compare the results of the SCM and the VCA for a $GaAs_{x}P_{1-x}$ bulk mixed-crystal. For SCM a super-cell with  $4\times4\times4$ bulk fcc elementary cells is used. The elementary cell for VCA is the same fcc-cell as for $GaP$ and $GaAs$, but with two virtual, weighted anions. The initial (i.e. before relaxation) lattice constant for a $GaAs_{x}P_{1-x}$ crystal is taken from Vegard's law\cite{denton1991vegard,nakamura2020crystal} using $a_{GaAs}=5.765 \, \AA$ and $a_{GaP}=5.53 \, \AA$.
In both cases the lattice constant has not been changed after relaxation. For SCM we use ten random configurations for each $x$-value (except for $x=0/1$) to get reasonable averages.   
\\
Secondly, to determine the stability of a particular surface reconstruction, we calculated the  surface-formation energy $\Omega$ per unit area \cite{bechstedt2012principles,ruiz2021inp}:
\begin{eqnarray}
\Omega =\frac{1}{A_{s}} \left[ E_{tot}-\sum_{i}N_{i}\mu_{i}\right].\label{eq:omega}
\end{eqnarray}
Here, $E_{tot}$  is the total energy of the slab, $A_{s}$ is the area of the surface unit cell, $N_{i}$ is the number of atoms of the species $i=A,B,C,...$ in the slab, and $\mu_{\textit{i}}$ is the chemical potential of species $i$.
Using the chemical potential $\mu_{\textit{i}}^{bulk}$ of the corresponding bulk crystal of species $i$ and the relation $\Delta\mu_{i}=\mu_{i}-\mu_{i}^{\textit{bulk}}$ this can be rewritten as:
\begin{eqnarray}
\Omega =\frac{1}{A_{s}} \left[ E_{tot}-\sum_{i}(\vartriangle\mu_{i}+\mu_{i}^{bulk})\right].\label{eq:omega1}
\end{eqnarray}
\\
Introducing the heat of formation for a binary ($A,B$) crystal $H_{AB}^{f} $:
\begin{eqnarray}
 H_{AB}^{f}=\mu_{A}^{bulk}+ \mu_{B}^{bulk}-\mu_{AB}^{bulk},
\end{eqnarray}
one can note the limits for $\Delta\mu_{i}$ imposed by the thermodynamic equilibrium:
\begin{eqnarray}
-H_{AB}^{f}\leqslant \Delta\mu_{i}  \leqslant 0.
\end{eqnarray}
The chemical potential of the mixed-crystal $GaAs_{x}P_{1-x}$ can be defined as:
\begin{eqnarray}
\mu_{GaAs_{x}P_{1-x}}^{bulk}= \mu_{\textrm{Ga}}+x\mu_{\textrm{As}}+(1-x)\mu_{\textrm{P}} ,
\label{eq:3}
\end{eqnarray}
 if one assumes, that the gas-phase is in equilibrium with the bulk and that the share $x$ of $As$ atoms in the gas-flow is the same as in the growing bulk.
\\
Then $H_{GaAs_{x}P_{1-x}}^{f}$ reads :
\begin{eqnarray}
H_{GaAs_{x}P_{1-x}}^{f} & = & \mu_{\textrm{Ga}}^{bulk}+x\mu_{\textrm{As}}^{bulk}+(1-x)\mu_{\textrm{P}}^{bulk}-\mu_{GaAs_{x}P_{1-x}}^{bulk},\label{eq:HF}
\end{eqnarray}
and eq. (\ref{eq:omega1}) extends to:
 \begin{eqnarray}
\Omega & = &\frac{1}{A_{s}} \bigg[ E_{tot}-N_{\textrm{Ga}}(\vartriangle\mu_{\textrm{Ga}}+\mu_{\textrm{Ga}}^{bulk})-N_{\textrm{H}}(\vartriangle\mu_{\textrm{H}}+\frac{1}{2}E_{H_{2}})\label{eq:HminusE-2-1-1}\\
 &  & -N_{\textrm{As}}\big(-\vartriangle\mu_{\textrm{Ga}}-H_{GaAs_{x}P_{1-x}}^{f}+(1-x)\,(\vartriangle\mu_{\textrm{As}}-\vartriangle\nonumber\mu_{\textrm{P}})+\mu_{\textrm{As}}^{bulk}\big)\\
 &  & -N_{\textrm{P}}\big(-\vartriangle\mu_{\textrm{Ga}}-H_{GaAs_{x}P_{1-x}}^{f}-x\,(\vartriangle\mu_{\textrm{As}}-\vartriangle\mu_{\textrm{P}})+\mu_{\textrm{P}}^{bulk}) \bigg]\nonumber
\end{eqnarray}
Differentiating eq. (\ref{eq:3}) with respect to $x$ and using the $\Delta\mu_{i}$ one finds:
\begin{eqnarray}
\vartriangle\mu_{\textrm{As}}-\vartriangle\mu_{\textrm{P}} & = & \frac{\partial \mu_{GaAs_{x}P_{1-x}}^{bulk}}{\partial x}-\mu_{\textrm{As}}^{bulk}+\mu_{\textrm{P}}^{bulk},
\end{eqnarray}
which we substitute in eq. (\ref{eq:HminusE-2-1-1}).
\\
The values of $\mu_{Ga}^{bulk}$, $\mu_{As}^{bulk}$, $\mu_{P}^{bulk}$, and $E_{H_{2}}$ have been calculated using VASP relative to the vacuum level as:  -2.9 eV, -4.68 eV, -5,18 eV, and -6.77 eV. These are close to those reported by Moll et al. \cite{moll1996gaas} and Kirklin et al. \cite{kirklin2015open}.
The differences of the chemical potentials $\vartriangle\mu_{\textrm{Ga}}=-0.25$ eV and $\vartriangle\mu_{\textrm{H}}=-0.55$ eV are chosen in such a way that the experimentally confirmed structures are obtained for the limiting cases ($x=0$: $(2 \times 2)$-2D-2H,\,  $x=1$: $\beta2(2\times4)$). 
\\
In our set up we use a slab with 15 atomic layers with initial fcc-coordination. The uppermost nine layers were allowed to relax while the others were fixed at their bulk positions. The lower slab surface, consisting of anions, has been passivated by pseudo-hydrogen with $Z = 0.75$. The vacuum has an extension of about four lattice constants. 
The calculations were performed with VASP within the generalized gradient approximation (GGA) and with the Perdew-Burke-Ernzerhof (PBE) functional \cite{perdew1996generalized,kobayashi2022machine}. 
The cut-off energy for the plane wave
basis is chosen to be 400 eV. An automatically generated k-point set with six points for each lateral dimension and one
point in vertical direction per $(2 \times 2)$ unit cell is used.
\section{RESULT and DISCUSSION}
\begin{figure}[htp]
    \centering
   \includegraphics[width=0.8\textwidth]{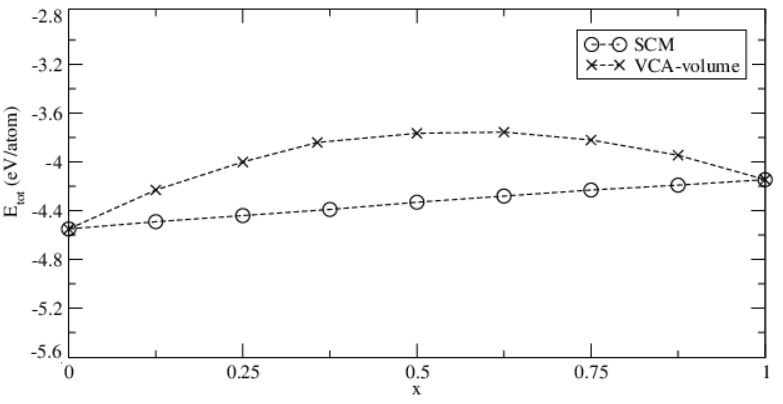}
   \caption{ Total energy (per atom) of a mixed-crystal $GaAs_{x}P_{1-x}$ vs. concentration $x$. The circles are the results given by the super-cell method and the $\times$-symbols are for the  Virtual Crystal Approximation. The differences increase for $x$ towards $\frac{1}{2}$ and vanish at the edges. The dashed lines are just to guide the eye.  }    \label{fig:mesh1}
\end{figure}

\begin{figure}[htp]
    \centering
   \includegraphics[width=0.8\textwidth]{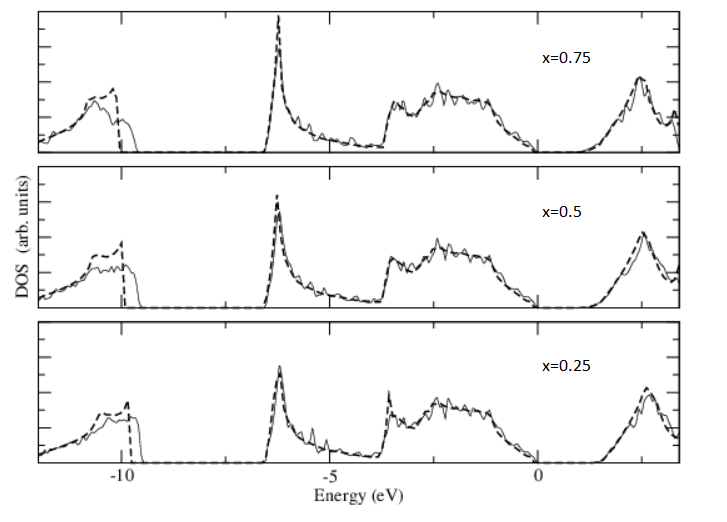}
   \caption{ Density of states of a mixed-crystal $GaAs_{x}P_{1-x}$ for different concentrations $x$. The gray solid line is for the super-cell method (SCM) and the black dashed lines for the Virtual Crystal Approximation (VCA). Besides the fact that neither method produces the exact solution -- VCA uses weighted virtual atoms, SCM suffers mainly from the periodic boundary conditions -- the over all similarity encourages us to consider the implementation of VCA as reliable.}    \label{fig:mesh2}
\end{figure}

\begin{figure}[htp]
    \centering
   \includegraphics[width=0.8\textwidth]{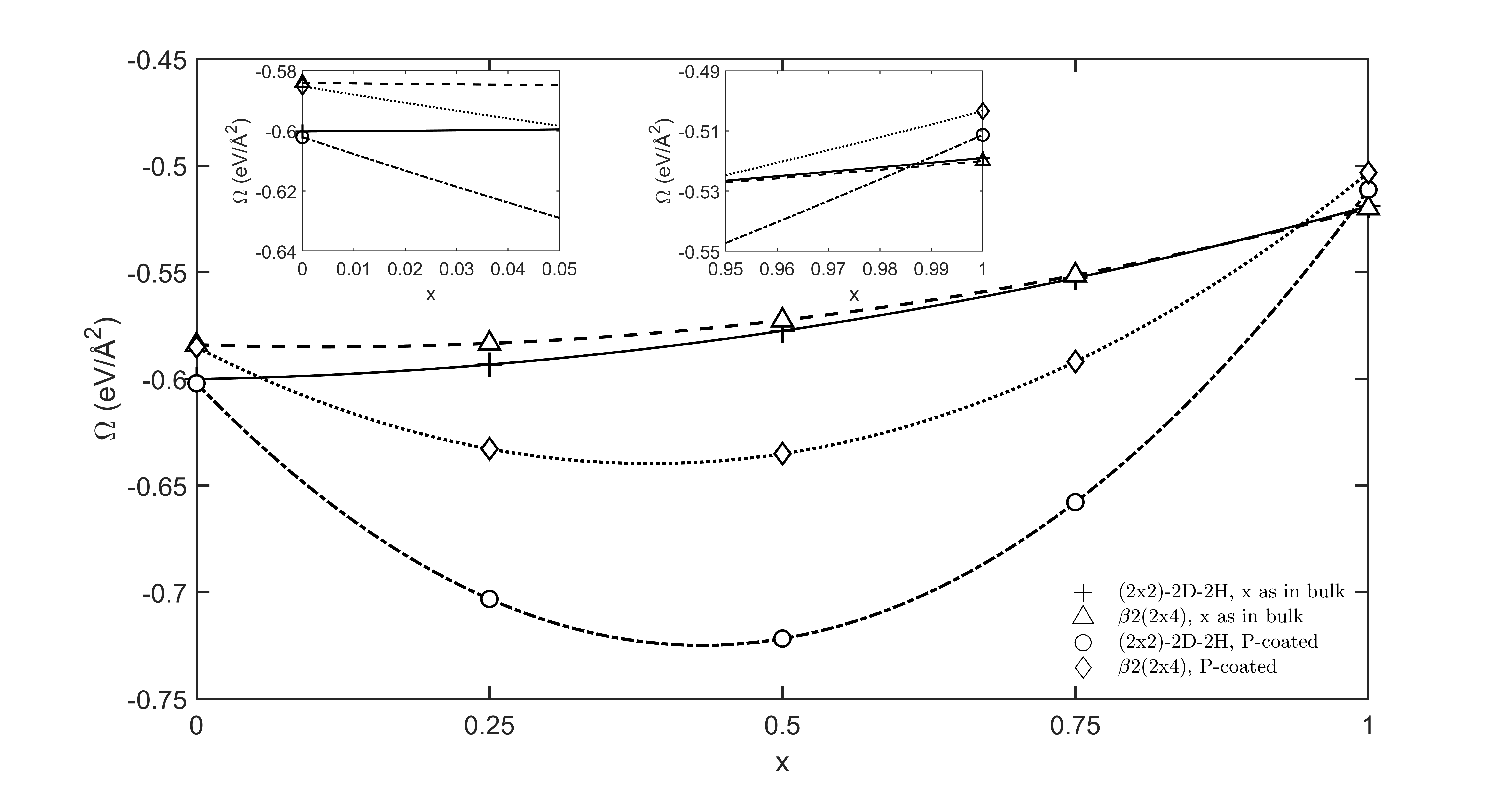}
   \caption{ Surface formation energy $\Omega$ versus concentration $x$, with $x$= 0, 0.25, 0.5, 0.75, 1. Circles and diamonds are for pure phosphorous surfaces  whereas +-symbols and triangles are for surfaces with the same $x$ as in the bulk. The lines are just to guide the eye. The insets display the situations at the edges of the $x$-interval in more detail. The (2$\times$2)-2D-2H $P$-coated geometry is over nearly the whole $x$-range energetically favored. The change to the $\beta2(2\times4)$ reconstruction seems to happen at rather large $x$.}    \label{fig:mesh3}
\end{figure}

\begin{figure}[htp]
    \centering
   \includegraphics[width=0.8\textwidth]{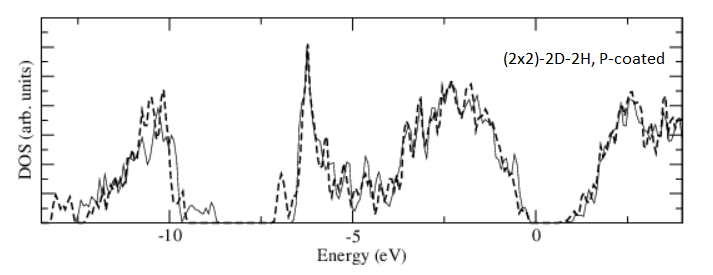}
   \caption{ Density of states (DOS) of (2$\times$2)-2D-2H with $x=0.5$ and only $P$ in the surface for a random configuration and the Virtual Crystal Approximation (VCA). The gray solid line is for the random configuration and the black dashed line is for the VCA. 
The result is that both curves have approximately the same shape. Due to the kind of comparison, one should not stress the differences. } \label{fig:mesh4}
\end{figure}

The implementation of the VCA method in VASP is  still under development (i.e. usable, but not optimized)\cite{jjdsds}. Therefore, for testing purpose, we use both methods, VCA and SCM, to calculate the total energy ($E_{tot}$) and density of states (DOS) for a bulk mixed-crystal $GaAs_xP_{1-x}$. 
\\
Firstly, we show in Fig. \ref{fig:mesh1} $E_{tot}$ (per atom) vs. the concentration $x$.  The circles are for the SCM and the $\times$-symbols for the VCA, the dashed lines are just to guide the eye. The result of the SCM shows a linear dependency with $x$, while the result of the VCA is non-linear.
We consider that the SCM depicts a more realist behaviour because it has real local interactions whereas the VCA handles a substitute system with virtual weighted anions \cite{yu2009atomistic}. Therefore, one expects that the deviations vanish at the edges of the $x$-interval and to be greatest in the middle. It should be noted, that $E_{tot}$ from the VCA has no buckles or discontinuities which is necessary when depending on a continuous parameter like the concentration $x$. On one hand this convinces us of a proper working implementation of the VCA, on the other it illustrates a known problem of the VCA, the overestimation of $E_{tot}$.
\\
In Fig. \ref{fig:mesh2} the DOS of $GaAs_xP_{1-x}$ for the SCM and the VCA are plotted for  $x=0.25/0.5/0.75$. The gray solid line is the result of the SCM and the black dashed one is for the VCA. The agreement is remarkable and it can be argued that the ad-hoc averaged potential creates energy landscape quite similar to that of the super-cell. The same has been reported by Sen and Ghosh for a different material system\cite{sen2016electronic}. This illustrates a strength of the VCA, it sketches the electronic states mostly correct.
\\
For the investigation of surfaces of $GaAs_xP_{1-x}$ mixed-crystals, we calculate $\Omega$ for four cases of group-$V$ terminated surfaces: $(2 \times 2)$-2D-2H and $\beta2(2\times4)$, with both a $P$ top-layer and with an  $As_xP_{1-x}$ top anion layer with the same concentration $x$ as in the underlying  $GaAs_xP_{1-x}$ bulk. In this work we are not considering slabs with an $As$ top-layer, since we are comparing our results with experimental findings where $P$-rich surfaces are grown.
In Fig. \ref{fig:mesh3} we show $\Omega$ vs. the concentration $x$, with $x$=  0, 0.25, 0.5, 0.75, 1. The circles and diamonds are for purely P-coated surfaces in the specified reconstructions whereas the +-symbols and the triangles are for surfaces with the same concentration $x$ as in the bulk (the lines are just to guide the eye). The curvature of $\Omega$ is somewhat surprising at first glance, because $E_{tot}$ is the leading term in eq. (\ref{eq:HminusE-2-1-1}). However, $\mu_{i}^{bulk}$, multiplied by the corresponding numbers of atoms, compensate for the overestimation of $E_{tot}$.
The insets give a better view on the situation at $x=0/1$. At $x=0$, pure $GaP$, the data points are degenerated and ordered: because the chemical potentials have been chosen so that the $(2 \times 2)$-2D-2H surface geometry is energetically favored over the $\beta2(2\times4)$ one. On the other edge, $x=1$, (pure $GaAs$), the preferred configuration is the $\beta2(2\times4)$ with an $As$ top-layer -- instead of the one with the $P$ layer. The transition between the two surfaces $(2 \times 2)$-2D-2H and $\beta2(2\times4)$ seems to occur at high concentrations of $As$, at least $x> 0.75$. 
The figure suggests that for nearly the whole $x$-range the preferred surface reconstruction is the $(2 \times 2)$-2D-2H type  with only phosphor in the surface, in conformance with the experimental findings \cite{zorn1994situ,supplie2020quantification}. Also, for the same top layer concentration $x$ as in the bulk the $(2 \times 2)$-2D-2H geometry is favored above the $\beta2(2\times4)$ up to $x=0.75$. All this indicates that the surface reconstruction depends mostly on the type of anions present on the surface and less on the composition of the bulk.
\\
In Fig. \ref{fig:mesh4} we compare the DOS of a $GaAs_{0.5}P_{0.5}$ mixed-crystal with a surface in a (2$\times$2)-2D-2H reconstruction  and only $P$ in the surface yielded by the VCA with that of a configuration with a random bulk. We restrict ourselves to the latter one because a SCM is not feasible, as mentioned. Though, the bulk of the random configuration is not a mixed-crystal -- strictly speaking, it is a valid configuration of a mixed-crystal -- it should give an idea of the DOS to be expected. The gray solid line is for the random configuration and the black dashed one for the VCA. The overall shape and position of the DOS are rather similar. Although one should not expect too much from such a comparison, the similarity of the DOS is decent and gives some confidence in the results of the VCA.
\subsection{SUMMARY}
Crystals with compositional disorder are frequently used in the manufacturing of compound semiconductor devices. Calculating the properties of such crystals is challenging. While we are using VASP, there are, in principle, two approaches to handle this problem: SCM and VCA.
On the one hand, the SCM is computational overwhelming and suffers from periodic boundary conditions, on the other hand, the VCA implementation is  not yet fully optimized and has an artificial interaction between virtual weighted anions on each site. We calculated $E_{tot}$ and DOS of bulk $GaAs_{x}P_{1-x}$. Both results reflect weak as well as strong points of the VCA: $E_{tot}$ is overestimated while the very good agreement of the two DOS illustrates, that the VCA gets the electronic structures mostly correct. To handle the composition $x$ on a surface we used an extended equation for $\Omega$. We compared two different surface reconstructions, $(2 \times 2)$-2D-2H and $\beta2(2\times4)$ with (i) only $P$-anions in the surface and (ii) the same concentration $x$ as in the  bulk vs. $x$. The chemical potentials, as free parameters, were chosen, so that the experimental findings at $x=0/1$ were reproduced.  We found that the $P$-coated $(2 \times 2)$-2D-2H reconstruction is energetically favored over a large range of the $x$-interval. This is in conformance with experimental reports, though it has to be mentioned that the finishing $P$-layer is produced on purpose, namely by changing temperature, pressure, and etc.. 
  \section*{ACKNOWLEDGEMENTS}
We thank Henning Schwanbeck and the University Computing Center of the TU Ilmenau for excellent working conditions and continuing support.
\bibliography{References}

    

\end{document}